\documentclass[%
 reprint,
 amsmath,amssymb,
 aps,
 prx,
 superscriptaddress,
]{revtex4-2}

\usepackage{graphicx}
\usepackage{dcolumn}
\usepackage{bm}
\usepackage{braket}
\usepackage{url}
\usepackage{multirow}
\usepackage{xcolor}
\usepackage[normalem]{ulem} 

\begin{document}
\title{Scattering Circular Dichroism from Spatially Modulated Anisotropic Magnetic Dipoles in BiFeO$_3$}

\author{Y. Ishii}
\email[Corresponding author: ]{ISHII.Yuta@nims.go.jp}
\affiliation{Center for Basic Research on Materials (CBRM), 
National Institute for Materials Science (NIMS), Tsukuba 305-0047, Japan}
\author{N. Sasabe}
\affiliation{Center for Basic Research on Materials (CBRM), 
National Institute for Materials Science (NIMS), Tsukuba 305-0047, Japan}
\author{T. Honda}
\affiliation{Institute of Materials Structure Science, High Energy Accelerator Research Organization, Tsukuba 305-0801, Japan}
\author{S. Kawachi}
\affiliation{Graduate School of Science, University of Hyogo, Koto, Hyogo 678-1297, Japan}
\author{T. Ito}
\affiliation{Electronics and Photonics Research Institute, National Institute of Advanced Industrial Science and Technology (AIST), Tsukuba, Ibaraki 305-8562, Japan}
\author{A. Sumiyoshiya}
\affiliation{General Incorporated Foundation Photon Science Innovation Center (PhoSIC), Sendai 980-0845, Japan}
\author{H. Nomura}
\affiliation{International Center for Synchrotron Radiation Innovation Smart, Tohoku University, Sendai 980-8577, Japan}
\author{T. Nakamura}
\affiliation{International Center for Synchrotron Radiation Innovation Smart, Tohoku University, Sendai 980-8577, Japan}
\affiliation{General Incorporated Foundation Photon Science Innovation Center (PhoSIC), Sendai 980-0845, Japan}
\author{T. Arima}
\affiliation{RIKEN Center for Emergent Matter Science (CEMS), Wako 351-0198, Japan}
\affiliation{Department of Advanced Materials Science, University of Tokyo, Kashiwa 277-8561, Japan}
\author{Y. Yamasaki}
\email[Corresponding author: ]{YAMASAKI.Yuichi@nims.go.jp}
\affiliation{Center for Basic Research on Materials (CBRM), National Institute for Materials Science (NIMS), Tsukuba 305-0047, Japan}
\affiliation{RIKEN Center for Emergent Matter Science (CEMS), Wako 351-0198, Japan}
\affiliation{International Center for Synchrotron Radiation Innovation Smart, Tohoku University, Sendai 980-8577, Japan}

\begin{abstract}
Antiferromagnets with broken time-reversal symmetry, including altermagnets, have attracted growing interest because they can exhibit ferromagnet-like responses, such as the anomalous Hall effect and magneto-optical effects, despite their compensated spin structures.
Such time-reversal symmetry breaking is characterized by ferroic time-reversal-odd multipoles, distinct from the conventional spin and orbital moments.
Among them, the anisotropic magnetic dipole (AMD), which reflects the local orbital anisotropy, shares the symmetry of magnetization and can have a ferroic component even when the spin moments cancel.
AMD responses have thus far been identified primarily in antiferromagnets with unit-cell magnetic periodicity, whereas spatially modulated AMDs in long-wavelength magnetic textures remain unexplored.
Here, we report resonant soft x-ray small-angle scattering measurements on the antiferromagnetic spin cycloid of the multiferroic BiFeO$_3$.
We observe pronounced circular dichroism in the resonant scattering intensity, i.e., scattering circular dichroism, which is not expected from the established cycloidal and Dzyaloshinskii-Moriya-interaction-induced sinusoidal spin structures.
We show that the staggered rotation of the FeO$_6$ octahedra gives rise to ferroic AMD components despite the antiparallel spin alignment.
As the local magnetic symmetry varies with the cycloidal phase, these AMDs are spatially modulated and produce the dichroism through interference with the sinusoidal spin component.
Our findings demonstrate that long-wavelength antiferromagnetic textures can host spatially modulated AMDs, opening new avenues for exploring hidden multipole textures in magnetic and multiferroic materials.
\end{abstract}

\maketitle

\section{Introduction}
Compensated spin order can coexist with pronounced magneto-optical and transport responses, making antiferromagnets a rich platform for exploring the interplay between magnetic symmetry and electronic structure~\cite{Smejkal2020_SciAdv,Smejkal2022_PRX,Smejkal2023_NatRevMater}.
These responses, such as the anomalous Hall and Nernst effects~\cite{Smejkal2020_SciAdv,Gonzalez2023,Reichlova2024,Sato2024,Takagi2025} and magneto-optical dichroism~\cite{Hariki2024,Amin2024,Hariki2025,Peter2025,Galindez2025}, require broken time-reversal symmetry, as in ferromagnets, and arise in a class of antiferromagnets, including altermagnets.
This time-reversal symmetry breaking is characterized by time-reversal-odd magnetic multipoles, which are not captured by the conventional spin and orbital moments.
In $\alpha$-MnTe and $\alpha$-Fe$_2$O$_3$, for example, the magnetic symmetry allows a ferroically ordered magnetic dipole despite the antiferromagnetic spin alignment~\cite{Yatsushiro2021,Cheong2023,Lovesey2023,sasabe2025,ishii2026}.
In the $d$-wave altermagnet MnF$_2$, by contrast, this role is played by ferroically ordered magnetic octupoles~\cite{Bhowal2024}.
Similarly, the large anomalous Hall effect in the noncollinear antiferromagnet Mn$_3$Sn has been discussed in terms of a ferroic order of cluster magnetic octupoles~\cite{Nakatsuji2015,Suzuki2017,XMCD_Kimata2021}.

Among these multipoles, the anisotropic magnetic dipole (AMD) is of particular importance, 
as it is a time-reversal-odd rank-1 axial tensor, like magnetization, and can carry the ferromagnet-like character of antiferromagnets~\cite{Hayami2021}.
The AMD is a spinful magnetic dipole described by the tensor product of a rank-2 quadrupole tensor $\bm{Q}^i$ and the spin operator $\bm{s}^i$, $\bm{t} = \sum_i \bm{Q}^i \bm{s}^i$~\cite{Oguchi2004,kusunose2020complete,hayami2024unified,YamasakiSTAM2025}.
Since the AMD depends on the anisotropy of the local electronic orbital states, antiferromagnetically coupled sublattices with distinct orbital states can host a ferroic AMD component, even when their spin moments are compensated.
Ferroic AMD responses have been observed in collinear and noncollinear antiferromagnets via x-ray magnetic circular dichroism (XMCD)~\cite{Yamasaki2020,XMCD_Kimata2021,sakamoto2021observation,sasabe2023ferroic,sasabe2025}, owing to its sensitivity to the AMD in addition to the spin and orbital moments~\cite{Thole1992,Carra1993,Chen1995,vanDerLaan2014,YamasakiSTAM2025}.

These hidden magnetic multipoles in antiferromagnets have thus far been observed predominantly in systems in which the periodicity of the magnetic order matches that of the crystallographic unit cell, in close analogy to ferromagnets.
However, these multipoles can also emerge in a wide class of long-wavelength antiferromagnetic orders, such as cycloidal or helical spin structures.
In such systems, the local magnetic symmetry varies with the spin configuration at each magnetic site, and the hidden multipoles can therefore be spatially modulated.
Because such modulated multipoles cancel upon averaging, they are inaccessible to spatially uniform probes such as XMCD, and antiferromagnetic states characterized by them have remained largely unexplored.
Exploring spatially modulated hidden multipoles thus paves the way for a new class of antiferromagnetic states beyond the uniform ferroic ordering of magnetic multipoles.

Here, we report resonant elastic x-ray scattering (REXS) measurements, which provide momentum-resolved access to such modulated multipoles, on the antiferromagnetic spin cycloid of the well-known multiferroic BiFeO$_3$.
We observe pronounced circular dichroism in the resonant scattering intensity at the magnetic satellite reflections.
This dichroism cannot be explained by the established cycloidal and Dzyaloshinskii-Moriya (DM)-interaction-induced sinusoidal spin structures, but arises from spatially modulated ferroic AMDs through interference with the sinusoidal spin component.

\begin{figure*}[tbp]
\begin{center}
\includegraphics[width=14cm]{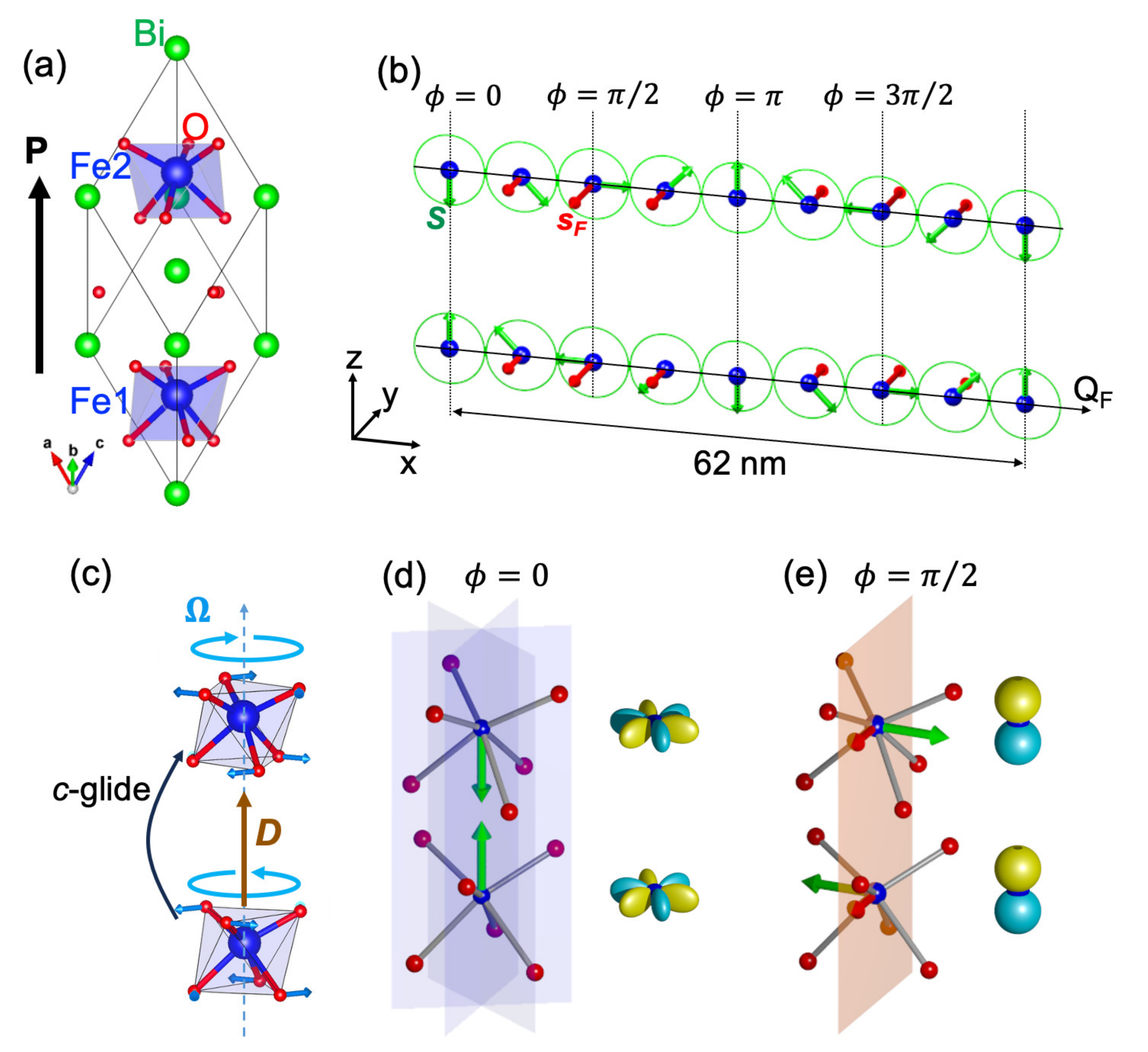}
\caption{
Local magnetic symmetry and magnetic multipoles in BiFeO$_3$.
(a) Trigonal crystal structure of BiFeO$_3$ (space group $R3c$) with the spontaneous ferroelectric polarization $\bm{P}$ along the [111]$_\text{rh}$ direction, visualized using VESTA~\cite{Momma_2011}.
The two Fe sites in the unit cell are labeled Fe1 and Fe2.
(b) Antiferromagnetic cycloidal structure with the modulation vector $\bm{Q}_{AF}$.
The spins are slightly canted by the Dzyaloshinskii-Moriya interaction, resulting in the sinusoidal structure with the modulation vector $\bm{Q}_{F}$.
Green and red arrows represent the antiferromagnetic cycloidal and sinusoidal spin moments ($\bm{S}$ and $\bm{s}_F$), respectively.
(c) Staggered rotational mode ($\bm{\Omega}$) of the FeO$_6$ octahedra around the [111]$_\text{rh}$ axis at the Fe1 and Fe2 sites.
Light blue arrows indicate the oxygen displacements associated with the rotation, and the brown arrow represents the DM vector $\bm{D}$ along the [111]$_\text{rh}$ axis induced by the $\bm{\Omega}$ rotation.
The Fe1 and Fe2 sites are related by a $c$-glide symmetry.
(d),(e) Spin configurations of the Fe1 and Fe2 sites (left panels) and the symmetry-allowed ferroic multipoles (right panels) at the cycloidal phases (d) $\phi = 0$ and (e) $\phi = \pi/2$.
Blue and orange planes in the left panels represent the $c$-glide and $c^\prime$-glide ($c$-glide combined with time reversal) planes, respectively.
Yellow and light blue lobes in the right panels represent the magnetic octupole at $\phi = 0$ and the magnetic dipole along the [111]$_\text{rh}$ direction at $\phi = \pi/2$, respectively.
}
\label{fig:structure}
\end{center}
\end{figure*}

\section{Local magnetic symmetry and magnetic multipoles in BiFeO$_3$}
\label{sec:symmetry}

Perovskite-type BiFeO$_3$ is a prototypical room-temperature multiferroic that exhibits one of the largest spontaneous electric polarizations $\bm{P}$ ($\sim$90~$\mu$C/cm$^2$), with a high Curie temperature ($\sim$1100~K) and antiferromagnetic order below a high N\'eel temperature ($\sim 653$~K)~\cite{Catalan2009}.
At room temperature, it has a rhombohedral crystal structure with the space group $R3c$ and two formula units in the unit cell, owing to a rotation of the FeO$_6$ octahedra around the $[111]_\text{rh}$ axis in the rhombohedral (rh) notation [Fig.~\ref{fig:structure}(a)]~\cite{michel1969atomic}.
Since the magnetic moments are coupled antiferromagnetically between nearest-neighbor ions, the magnetic structure is, to a first approximation, a $G$-type antiferromagnet with magnetic moments of $\mu_{\text{Fe}}=4.11~\mu_B$~\cite{fischer1980temperature,Lebeugle2008}.
$\bm{P}$ is parallel to the $[111]_\text{rh}$ axis [Fig.~\ref{fig:structure}(a)] and modifies the $G$-type order into a cycloidal magnetic order with a wavelength of 62~nm along the $[\bar{1}10]_\text{rh}$ direction [Fig.~\ref{fig:structure}(b)]~\cite{Sosnowska1982}.
A staggered rotation of the FeO$_6$ octahedra (referred to as the $\bm{\Omega}$ rotational mode) with a rotation angle of $\Omega = 13^\circ$ about the $[111]_\text{rh}$ axis occurs at the two Fe sites (Fe1 and Fe2) in the unit cell [Fig.~\ref{fig:structure}(c)]~\cite{Kubel1990}.
This rotation allows the DM interaction between the two Fe ions, with the DM vector $\bm{D}$ pointing along the $[111]_\text{rh}$ axis owing to the threefold rotational symmetry~\cite{Zvezdin_2012}.
The DM interaction cants the antiparallel moments when they are perpendicular to the $[111]_\text{rh}$ direction, but not when they are parallel to it~\cite{Ederer2005}.
The canting angle of $\alpha \sim 1.3^{\circ}$ determined by small-angle neutron scattering corresponds to a weak ferromagnetic moment of $|\bm{M}_{\text{Fe}}| \sim 0.09~\mu_B$/Fe~\cite{Ramazanoglu2011}.
The magnetic structure can thus be described as a combination of the cycloidal structure with the modulation vector $\bm{Q}_{AF}=(1-q_m,1+q_m,1)_\text{rh}$, where $q_m \sim 4.5\times 10^{-3}$ (r.l.u.), and the collinear sinusoidal weak ferromagnetic structure with $\bm{Q}_{F}=(-q_m,q_m,0)_\text{rh}$.
This coupling between the electric polarization and the magnetic order gives rise to large magnetoelectric effects, such as electric-field control of magnetization and magnetic-field control of ferroelectric polarization~\cite{Zhao2006,Lebeugle2008}.

Whether a ferroic magnetic multipole is symmetry allowed in an antiferromagnet is determined by the symmetry operation relating the antiferromagnetic sublattices.
In BiFeO$_3$, the two sublattice sites Fe1 and Fe2 are related by a $c$-glide symmetry operation [Fig.~\ref{fig:structure}(c)].
Since the N\'eel vector $\bm{N}$ rotates within the cycloidal plane, the local magnetic symmetry of the Fe1--Fe2 pair depends on the cycloidal phase $\phi$.
Figures~\ref{fig:structure}(d) and \ref{fig:structure}(e) show the spin configurations and the symmetry-allowed ferroic multipoles for the Fe1 and Fe2 sites at $\phi = 0$ and $\pi/2$, respectively.
At $\phi = 0$, $\bm{N}$ is oriented along the [111]$_\text{rh}$ direction, and the threefold rotational symmetry is preserved together with the three $c$-glide symmetries indicated by the blue planes.
These symmetries forbid a ferroic magnetic dipole, whereas a ferroic magnetic octupole is allowed.
In contrast, at $\phi = \pi/2$, $\bm{N}$ is oriented along the [$\bar{1}10$]$_\text{rh}$ direction, which breaks the threefold rotational symmetry and leaves a single $c^{\prime}$-glide symmetry ($c$-glide combined with time reversal).
This reduced symmetry allows a ferroic magnetic dipole along the [111]$_\text{rh}$ direction, orthogonal to the DM-induced canted spin component.

The ferroic magnetic dipole along the [111]$_\text{rh}$ direction at $\phi = \pi/2$ can be understood in terms of the AMD given by the product of the electric quadrupole of the local orbital state and the spin.
Taking the $z$ and $x$ axes along the [111]$_\text{rh}$ and [$\bar{1}10$]$_\text{rh}$ directions, respectively, the AMD component $t_z$ involves the product of the electric quadrupole component $Q_{zx}$ and the spin component $s_x$.
For $\bm{N}\parallel x$, the spin-orbit coupling breaks the threefold symmetry of the local orbital state and allows a $Q_{zx}$ component; the $\bm{\Omega}$ rotation, with opposite rotation angles at the Fe1 and Fe2 sites, makes this component staggered between the two sites.
Combined with the antiparallel spins, this staggered quadrupole component yields a $t_z$ contribution with the same sign at the Fe1 and Fe2 sites.
This symmetry-allowed dipole along the [111]$_\text{rh}$ direction can be represented as a ferroic AMD component, even though the spin moments at the two sites are antiparallel.
Consequently, the ferroic AMD is spatially modulated along the cycloid.

\begin{figure*}[tbp]
\begin{center}
\includegraphics[width=16cm]{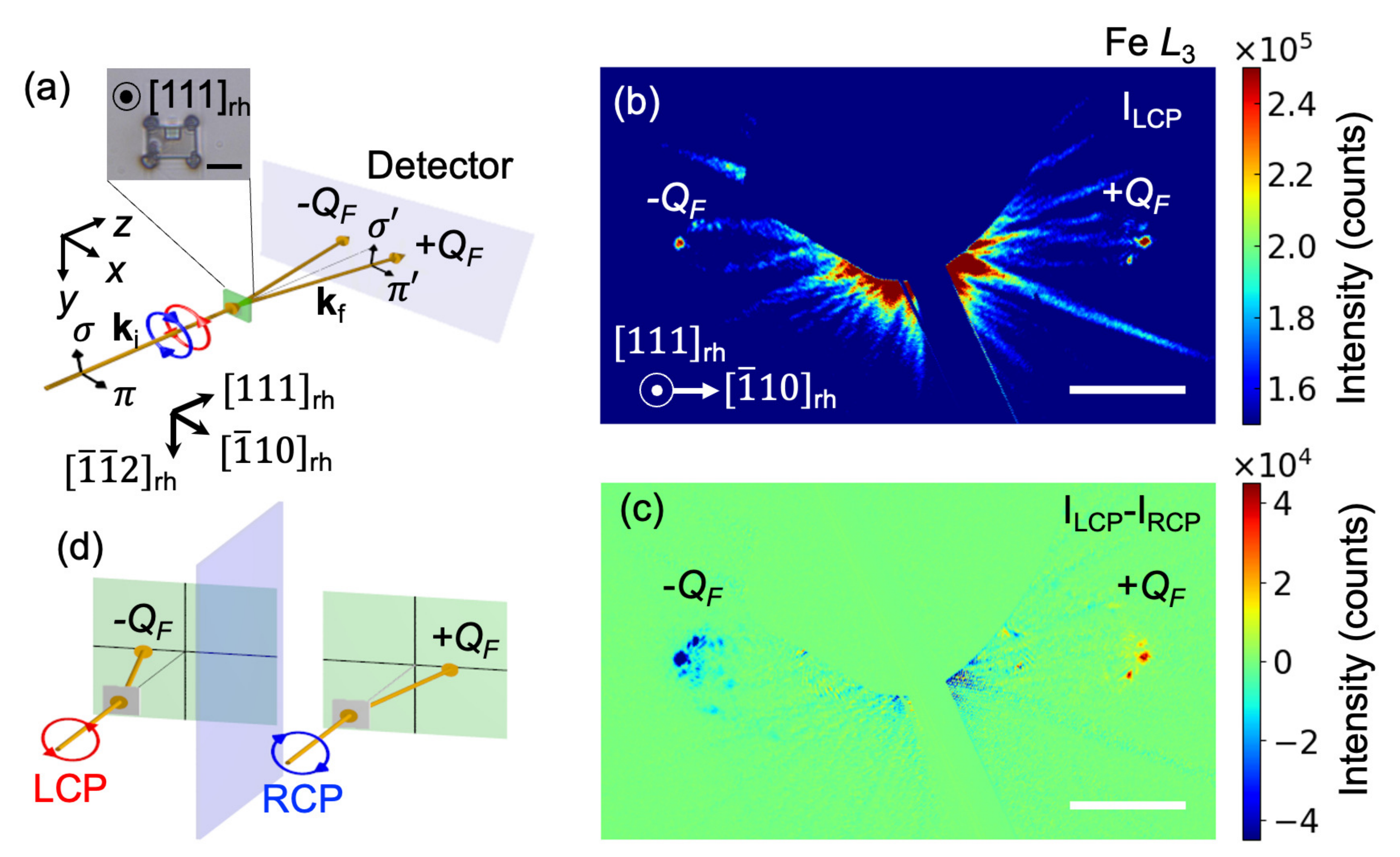}
\caption{
Circular dichroism in resonant soft x-ray scattering.
 (a) Schematic of the small-angle REXS setup.
 Left and right circularly polarized (LCP and RCP, respectively) soft x-rays are incident on the BiFeO$_3$ lamella, and the scattered intensity is recorded on a 2D detector.
Incident x-rays are parallel to the [111]$_{\rm rh}$ direction.
The inset shows a scanning electron microscope (SEM) image of the lamella fixed on a Si$_3$N$_4$ membrane coated with Au.
(b) REXS pattern measured at the Fe $L_3$ edge ($E$ = 709~eV) 
with LCP soft x-rays. 
The sample exhibits pronounced anisotropic magnetic scattering along the $\bm{Q}_F$ direction.
(c) Difference image between LCP and RCP x-rays, $\Delta I= I_\textrm{LCP}-I_\textrm{RCP}$, highlighting the circular dichroism in scattering intensity. 
The scale bars in (b) and (c) represent 0.05~nm$^{-1}$.
(d)
Configuration of REXS with mirror symmetry.
The blue plane represents the mirror plane.
The helicity of the circularly polarized soft x-rays is reversed under a mirror operation, leading to the intensity relation at $\pm \bm{Q}_F$,
$I_{\mathrm{LCP}}(\pm\bm{Q}_F) = I_{\mathrm{RCP}}(\mp\bm{Q}_F)$.
}
\label{fig:rexs}
\end{center}
\end{figure*}

Such a spatially modulated AMD cannot be detected by XMCD, which averages over the illuminated region.
REXS, by contrast, provides momentum-resolved information and is particularly powerful for investigating magnetic and electronic multipoles in solids.
In REXS, scattering occurs via resonant transitions between core and valence states and reflects the symmetry of the valence electronic configuration.
Multipoles that preserve inversion symmetry, such as electric charge, magnetic dipole, and electric quadrupole, are accessible through parity-even electric dipole-dipole transitions.
Different multipolar contributions can be distinguished by the polarization dependence of the scattered x-rays, and interference between them can reveal otherwise hidden orders; for example, interference scattering with circularly polarized x-rays has enabled the detection of orbital currents in CuO~\cite{Scagnoli2011}.
REXS thus extends XMCD into reciprocal space and is ideally suited for probing the spatially modulated AMD expected in BiFeO$_3$.

\begin{figure*}[tbp]
\begin{center}
\includegraphics[width=16cm]{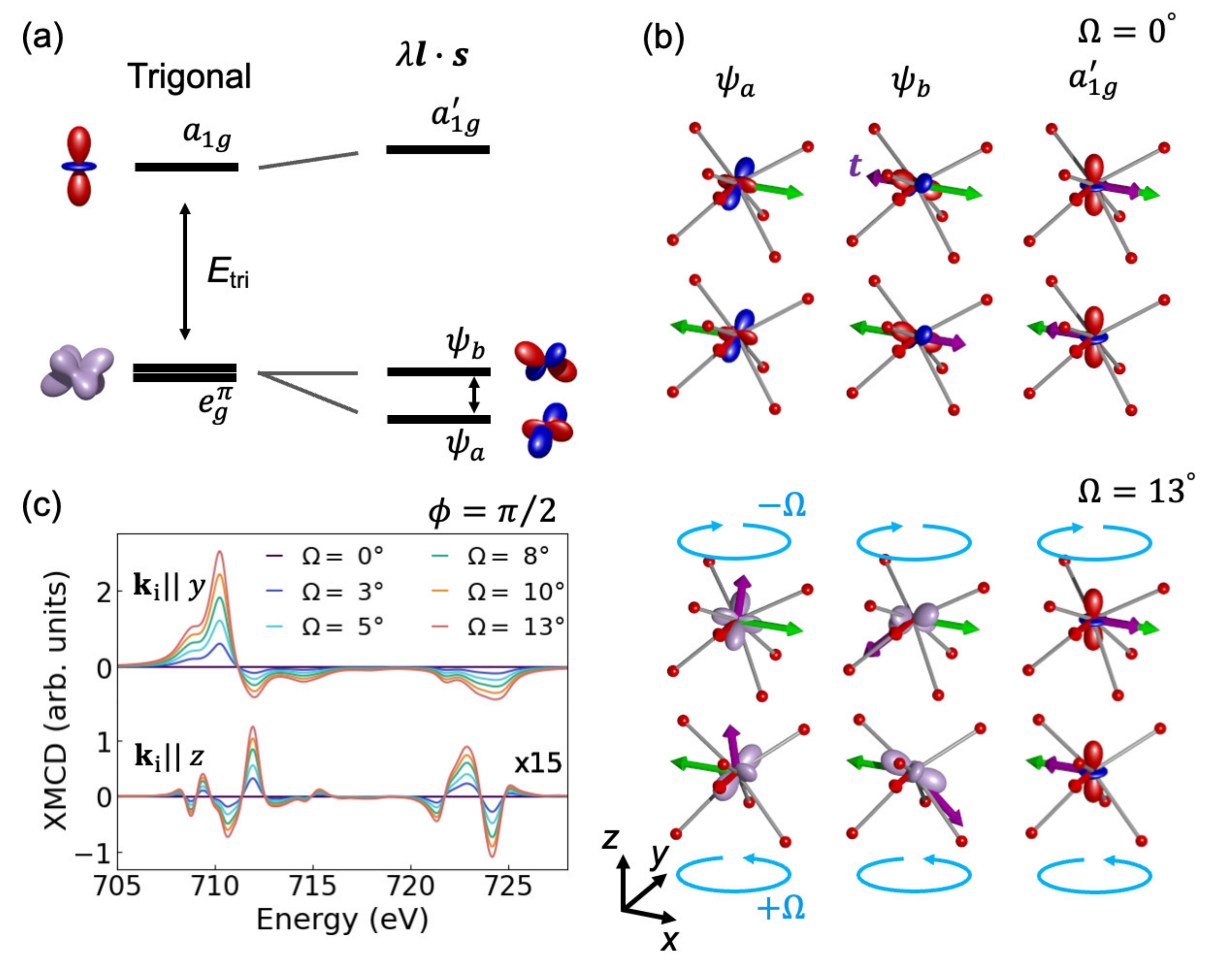}
\caption{
Ferroic AMD component induced by the $\bm{\Omega}$ rotation.
(a) Electronic states under the trigonal crystal field $E_{\rm tri}$ and the spin-orbit interaction $H_{LS} = \lambda {\bm l}\cdot{\bm s}$ at the cycloidal phase $\phi = \pi/2$.
For simplicity, only the $t_{2g}$ orbitals ($a_{1g}$ and $e_g^{\pi}$) are considered.
The weak $H_{LS}$ splits the $e_g^{\pi}$ states into $\psi_a$ and $\psi_b$ and modifies the $a_{1g}$ state into the $a_{1g}^{\prime}$ state.
(b) Spin and orbital configurations of the $\psi_{a,b}$ and $a_{1g}^{\prime}$ states at $\phi = \pi/2$ for $\Omega = 0^{\circ}$ and $13^{\circ}$, calculated by exact diagonalization.
Red, blue, and purple lobes represent the orbitals of the $\psi_{a,b}$ and $a_{1g}^{\prime}$ states.
Green, red, and purple arrows represent the antiferromagnetic spin moments ($\bm{S}$), the canted ferroic spin component along the $y$ axis ($\bm{s}_F$), and the expectation values of the AMD ($\bm{t}$) in the intermediate state, respectively.
(c) Calculated XMCD spectra along the $y$ (upper panel) and $z$ (lower panel) axes for various $\Omega$, arising mainly from $\bm{s}_F$ and the ferroic AMD, respectively.
}
\label{fig:amd}
\end{center}
\end{figure*}

\begin{figure*}[tbp]
\begin{center}
\includegraphics[width=17.5cm]{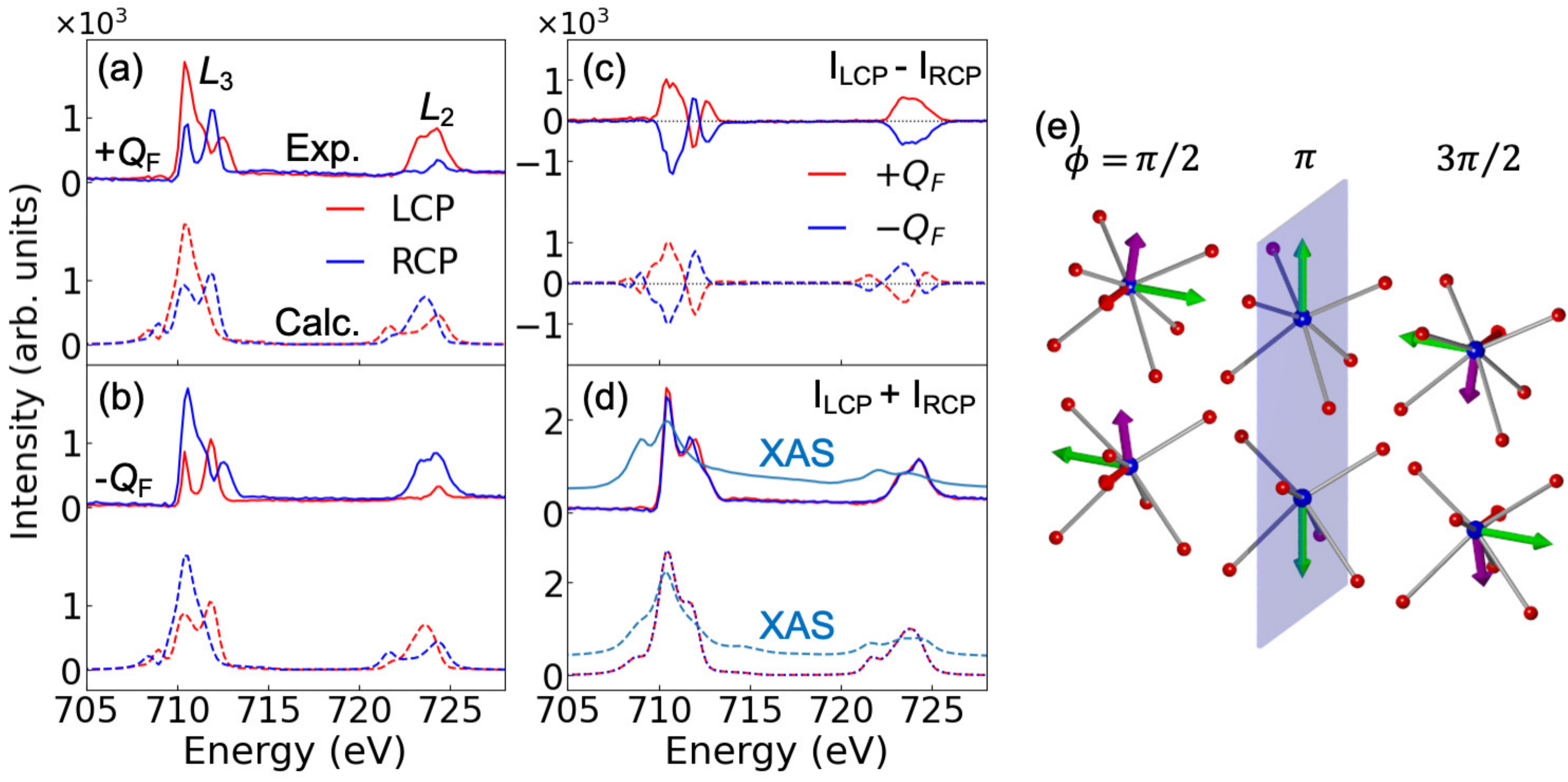}
\caption{
Experimental and theoretical REXS energy spectra.
(a),(b) Energy spectra of 
REXS near the Fe $L_{2,3}$ edges
obtained at (a) $\bm{Q}=+\bm{Q}_F$ 
and (b) $\bm{Q}=- \bm{Q}_F$ with 
left and right circularly polarized (LCP and RCP)
incident x-rays.
Solid and dashed lines 
represent the experimental and 
calculated spectra, respectively.
(c),(d) Energy spectra of (c) the difference and (d) the sum of the REXS intensities for LCP and RCP
($\Delta I = I_\textrm{LCP}-I_\textrm{RCP}$ and $\Sigma I = I_\textrm{LCP}+I_\textrm{RCP}$)
acquired at $\bm{Q}=\pm \bm{Q}_{F}$, 
together with the x-ray absorption spectra (XAS).
(e) AMD configurations of the $\psi_a$ states at the cycloidal phases $\phi = \pi/2$ and $3\pi/2$.
The configurations at these phases are related by the $c$-glide symmetry located at $\phi = \pi$, represented by the blue plane, which leads to the intensity relation $I_{\mathrm{LCP}}(\pm\bm{Q}_F) = I_{\mathrm{RCP}}(\mp\bm{Q}_F)$.
}
\label{fig:spectra}
\end{center}
\end{figure*}

\section{Experimental details}
BiFeO$_3$ single crystals were grown using the laser floating zone (LFZ) method~\cite{Ito2011}. 
The crystal axis directions were determined from Laue x-ray diffraction.
The crystal quality of the present sample, along with its dielectric and magnetic properties, has been investigated in previous studies~\cite{Ito2011,Kawachi2017,Kawachi2019}.
Samples suitable for soft x-ray small-angle scattering 
were prepared 
by focused ion beam (FIB) milling. 
A region of interest with the desired crystallographic orientation was selected from the bulk single crystal. 
Coarse milling was  performed using a Ga ion beam to isolate a lamella, which was subsequently lifted out using a micromanipulator and mounted onto a Si$_3$N$_4$ membrane support. 
The lamella was thinned step by step to the desired thickness, followed by final polishing using low-kV and low-current ion milling to minimize surface damage and redeposition. 

Small-angle REXS measurements 
were carried out at the soft x-ray beamlines BL-16A, Photon Factory, KEK, Tsukuba, Japan and BL14U, NanoTerasu, Sendai, Japan,
using soft x-ray diffractometers~\cite{Yamasaki2013,Ishii2025}
in a high-vacuum chamber with a base pressure of $1\times 10^{-8}$~Torr.
All measurements were performed at room temperature.
The incident soft x-rays 
were tuned to left- and right-circular polarizations. 
A soft x-ray CCD camera or CMOS camera positioned downstream of the sample was used to record the REXS intensity.
All XAS and REXS spectra were obtained by measuring the transmitted or scattered x-rays while scanning the incident photon energy.

\section{Circular dichroism in resonant x-ray scattering}

The geometry of the small-angle REXS
measurement is shown in Fig.~\ref{fig:rexs}(a).
The data are obtained around the Fe $L_3$ and $L_2$ edges
with left- and right-circularly polarized (LCP and RCP) incident x-rays.
The incident x-rays
are parallel to the [111]$_\text{rh}$
direction, which is parallel to the ferroelectric polarization.
The incident soft x-rays are transmitted through the lamella and lead to magnetic scattering signals in the small-angle region at positions corresponding to the characteristic wave vector $\bm{Q}_F$.
Spotlike magnetic diffraction peaks 
are observed at $\bm{Q} = \pm \bm{Q}_F$ 
along the [$\bar{1}10$]$_\text{rh}$ direction
with twofold symmetry,
as shown in Fig.~\ref{fig:rexs}(b).
The $R3c$ space group
generates  
three magnetic domains due to the threefold symmetry,
resulting in sixfold magnetic scattering, 
as reported in previous small-angle neutron scattering experiments~\cite{Ramazanoglu2011}.
We attribute the twofold magnetic scattering to cycloidal-domain selection by the uniaxial strain induced during sample preparation (see Appendix~\ref{app:strain}), as reported by neutron scattering under uniaxial strain~\cite{Lee2008} and in epitaxial BiFeO$_3$ thin films, where uniaxial strain is effectively imposed along a direction orthogonal to the spin-cycloidal propagation vector~\cite{haykal2020,Dufour2023}.

Figure~\ref{fig:rexs}(c) 
presents 
the difference image of the REXS intensity 
between LCP and RCP
incident x-rays, 
$\Delta I = I_\textrm{LCP}-I_\textrm{RCP}$.
Clear scattering circular dichroism  is observed at the magnetic scattering vector $\bm{Q}_F$
along the [$\bar{1}10$]$_\text{rh}$ direction.
Furthermore, the dichroism reverses in sign between $+\bm{Q}_F$ and $-\bm{Q}_F$, with a corresponding inversion in intensity:
$I_\textrm{LCP}(\pm\bm{Q}_F) = I_\textrm{RCP}(\mp\bm{Q}_F)$.
This relation is preserved at all measured photon energies, as will be discussed below with reference to the energy-resolved REXS spectra at $\pm\bm{Q}_F$ shown in Figs.~\ref{fig:spectra}(a) and \ref{fig:spectra}(b).
This antisymmetric behavior suggests that the magnetic configuration in the sample retains mirror symmetry.
Specifically, the observed reversal 
implies 
that the scattering process at $+\bm{Q}_F$ with RCP x-rays is equivalent to that at $-\bm{Q}_F$ with LCP x-rays [Fig.~\ref{fig:rexs}(d)].

Such circular dichroism in resonant x-ray scattering has been reported in materials exhibiting cycloidal spin structures~\cite{Fabrizi2009,Johnson2013}, where it originates from the chirality of the cycloid and appears with opposite signs at the magnetic satellite reflections, depending on the population of chirality domains.
However, this mechanism can be excluded in the present case, because magnetic scattering from the antiferromagnetic cycloidal component with the propagation vector $\bm{Q}_{AF}$ is forbidden at $\pm\bm{Q}_F = \pm(-q_m, q_m, 0)_\text{rh}$ by the $c$-glide symmetry of the crystal (see Sec.~I of the Supplemental Material~\cite{SM}).

Another possible contribution is magnetic scattering originating from the sinusoidal spin structure, which is modulated with $\bm{Q}_F$ and therefore contributes to the intensity at $\pm\bm{Q}_F$.
The polarization dependence of the resonant magnetic scattering factor is given by $f^j_{\varepsilon^\prime\varepsilon} \propto \hat{\bm{m}}_j \cdot (\bm{\varepsilon}^\prime\times \bm{\varepsilon})$, where $\hat{\bm{m}}_j$ is the quantization axis of the magnetic dipole at the $j$th site, and $\bm{\varepsilon}=\sigma, \pi$ and $\bm{\varepsilon^\prime}=\sigma^\prime,\pi^\prime$ are the polarizations of the incident and scattered x-rays, respectively [see Fig.~\ref{fig:rexs}(a)].
The circular dichroism in the REXS intensity is derived as~\cite{Lovesey_2008}
\begin{equation}
\Delta I\propto \text{Im}(F_{\pi^\prime\sigma}F_{\pi^\prime\pi}^*),
\label{eq:CD}
\end{equation}
where $F_{\varepsilon^\prime\varepsilon} =\sum_j f^{j}_{\varepsilon^\prime\varepsilon} (E)e^{i\bm{q}\cdot\bm{r}_j}$ is the structure factor.
For the sinusoidal spin structure, only the $F_{\pi^\prime\pi}$ component remains finite, whereas all other components vanish, resulting in the absence of circular dichroism [Eq.~(\ref{eq:Fcomp}) in Appendix~\ref{app:formalism}].
Furthermore, the reduced local magnetic symmetry for $\bm{N} \parallel [\bar{1}10]_\text{rh}$ also allows an extremely small ferromagnetic spin component along the [111]$_\text{rh}$ direction, giving rise to a tilted sinusoidal spin structure, and this component contributes to the $F_{\pi^\prime\sigma}$ term.
However, the tilted sinusoidal spin structure likewise does not generate circular dichroism, because the $F_{\pi^\prime\sigma}$ and $F_{\pi^\prime\pi}$ terms, both originating from the spin moments, exhibit nearly identical energy spectra, and the imaginary part of $F_{\pi^\prime\sigma}F_{\pi^\prime\pi}^*$ vanishes.
Therefore, neither the cycloidal nor the sinusoidal spin component can account for the observed circular dichroism.
This indicates the presence of an additional magnetic component that is hidden in the established spin structure and is modulated with the wave vector $\bm{Q}_F$, which we identify as the ferroic AMD in the following section.

\section{Ferroic AMD component in antiferromagnetic sublattices}
To uncover 
the origin of the dichroic magnetic scattering, 
we examine the $3d$ electronic states at the cycloidal phases $\phi = \pi/2$ and $3\pi/2$, where the magnetic dipole along the [111]$_\text{rh}$ direction is symmetry allowed.
Here, in the intermediate state of the REXS process, we consider the down-spin 
$3d$ states 
because of the fully occupied up-spin electronic states of the Fe$^{3+}$ ion 
and the weakness of spin-orbit coupling.
Under a trigonal crystal field, the $t_{2g}$ states are split into $a_{1g}$ and $e_g^{\pi}$, as illustrated in Fig.~\ref{fig:amd}(a) (see also Sec.~II of the Supplemental Material~\cite{SM}).
The doubly degenerate $e_g^{\pi}$ states are further split into two different states (hereafter denoted as $\psi_a$ and $\psi_b$) by the spin-orbit interaction $H_{LS}$, and the $a_{1g}$ state is correspondingly modified into the $a_{1g}^{\prime}$ state.

Figure~\ref{fig:amd}(b) displays the spin and orbital configurations of the $\psi_a$, $\psi_b$, and $a_{1g}^{\prime}$ states for $\Omega = 0^{\circ}$ and $13^{\circ}$, obtained by exact diagonalization within a single-electron picture, with the crystal field derived from the crystal structure (see Sec. II of the Supplemental Material).
For simplicity, the calculation is performed in the limit of weak spin-orbit coupling.
Green and red arrows represent the antiferromagnetic spin moments ($\bm{S}$) and the canted ferroic spin component ($\bm{s}_F$), respectively, and purple arrows represent the calculated expectation values of the AMD ($\bm{t}$).
In the absence of the $\bm{\Omega}$ rotation, the AMD moments are oriented parallel or antiparallel to the antiferromagnetic spin direction.
Consequently, the AMD adopts an antiferroic arrangement at the Fe1 and Fe2 sites and yields no contribution to the REXS intensity at $\pm\bm{Q}_F$.
The $\bm{\Omega}$ rotation induces a $z$ component of the AMD ($t_z$) for the $\psi_a$ and $\psi_b$ states with the same sign at the Fe1 and Fe2 sites, confirming the ferroic AMD expected from the symmetry analysis in Sec.~\ref{sec:symmetry}.
This ferroic $t_z$ is orthogonal to the canted spin component $\bm{s}_F$ along the $y$ axis.
The value of $t_z$ varies periodically with a period of 120$^{\circ}$ as a function of the rotation angle, reflecting the threefold symmetry (Fig.~\ref{fig:omega} in Appendix~\ref{app:omega}).
In contrast, the $\bm{\Omega}$ rotation does not modify the AMD in the $a_{1g}^{\prime}$ state, since this state is rotationally symmetric about the $z$ axis.
The spin, orbital, and AMD configurations at $\phi = 0$, $\pi/2$, $\pi$, and $3\pi/2$ are shown in Fig.~\ref{fig:fxyz}(a) in Appendix~\ref{app:cycloid}.

Since REXS is a resonant process, it probes the intermediate states, specifically, the transitions from $2p^63d^n$ to $2p^53d^{n+1}$ at the $L$ edge of $3d$ transition metals.
The ferroic AMD at $\phi = \pi/2$ and $3\pi/2$ is allowed by the local magnetic symmetry of the ground-state magnetic configuration, as discussed in Sec.~\ref{sec:symmetry}.
In the ground state, however, its magnitude is expected to be small owing to the nearly isotropic $3d^5$ configuration of Fe$^{3+}$.
In the intermediate states, the additional $3d$ electron and the core hole enhance the orbital anisotropy, so that the symmetry-allowed AMD contribution is amplified in the resonant response.
Because the resonant x-ray response remains governed by the local orbital anisotropy and the underlying crystal and magnetic symmetry~\cite{Hannon1988,Haverkort2010}, the emergence of the ferroic AMD component is determined by the octahedral rotation, and its spatial modulation by the local magnetic symmetry varying along the cycloid.

Figure~\ref{fig:amd}(c) presents the calculated XMCD spectra along the $y$ and $z$ axes for various $\Omega$ angles for the spin configuration of the Fe1 and Fe2 clusters at $\phi = \pi/2$, where the spectra are obtained as the sum of the contributions from the Fe1 and Fe2 sites.
The calculations were performed using the \texttt{EDRIXS} package~\cite{WANG2019}, with molecular fields and crystal fields that reproduce the local magnetic and crystal structures at each site (see Appendix~\ref{app:spectra} for details).
The ferroic components $\bm{s}_F$ and $t_z$ give rise to XMCD signals along the $y$ and $z$ axes, respectively.
The canted spin component produces a conventional XMCD spectrum resembling those of ferromagnets, whereas the XMCD along the $z$ axis originating from the AMD exhibits an oscillatory spectral shape, similar to that reported for $\alpha$-MnTe and $\alpha$-Fe$_2$O$_3$~\cite{Hariki2024,ishii2026}.
These distinct spectral shapes of the $\bm{s}_F$ and AMD contributions are essential for the interference discussed below.

The REXS measurements at $\bm{Q}=\pm \bm{Q}_F$ probe the spatial modulation of this ferroic AMD (Appendix~\ref{app:formalism}).
The DM-induced sinusoidal spin component $\bm{s}_F$ and the AMD moment contribute to different polarization channels of the resonant scattering amplitude. 
In the present geometry, $\bm{s}_F$ along the $y$ axis mainly contributes to the $F_{\pi'\pi}$ channel, 
whereas the AMD moment along the $z$ axis gives rise to the $F_{\pi'\sigma}$ channel.
Since these two contributions originate from different scattering processes, their scattering factors exhibit different energy dependences in both amplitude and phase [Fig.~\ref{fig:fxyz}(b) in Appendix~\ref{app:cycloid}].
Their interference, represented by Eq.~(\ref{eq:CD}), therefore produces finite circular dichroism.
This circular-dichroic term is proportional to $\cos 2\theta \sin 2\theta$ [Eq.~(\ref{eq:CDfull}) in Appendix~\ref{app:formalism}], because the polarization dependences of $F_{\pi'\sigma}$ and $F_{\pi'\pi}$ are given as $F_{\pi'\sigma} \propto \cos2\theta$ and $F_{\pi'\pi} \propto \sin2\theta$, respectively [Eq.~(\ref{eq:Fcomp})], where $2\theta$ denotes the scattering angle.
Under the transformation from $+\bm{Q}_F$ to $-\bm{Q}_F$, the scattering direction is reversed and hence $2\theta \rightarrow -2\theta$, giving $\Delta I(+\bm{Q}_F)=-\Delta I(-\bm{Q}_F)$.

The circular-dichroic response is dominated by the $\phi=\pi/2$ and $3\pi/2$ phases, where the AMD acquires a ferroic $z$-axis component and interferes with the sinusoidal spin component.
By contrast, at $\phi = 0$ and $\pi$, the AMD is oriented parallel to the spin, resulting in an antiferroic alignment, as shown in Fig.~\ref{fig:fxyz}(a).
This is consistent with the local magnetic symmetry at these phases, which forbids a ferroic magnetic dipole.
Consequently, the $\phi = 0$ and $\pi$ phases do not contribute to the circular dichroism.

\section{Energy spectra of resonant x-ray scattering}
To examine whether the observed dichroism is also present for other intermediate states, the energy-resolved scattering spectra at $\pm \bm{Q}_F$ around the Fe $L_{2,3}$ edges are presented in Figs.~\ref{fig:spectra}(a) and \ref{fig:spectra}(b).
Notably, the energy spectra of the REXS intensity for LCP and RCP are clearly different.
Furthermore, the spectra for LCP and RCP are nearly interchanged between the $\pm \bm{Q}_F$ reflections, indicating the relation $I_\textrm{LCP}(\pm \bm{Q}_{F})=I_\textrm{RCP}(\mp \bm{Q}_{F})$.
Figures~\ref{fig:spectra}(c) and \ref{fig:spectra}(d) show the difference spectra ($\Delta I$) and the sum spectra ($\Sigma I$) at $\pm \bm{Q}_F$, respectively, together with the x-ray absorption spectra (XAS).
These results confirm the relations $\Sigma I (+\bm{Q}_F)=\Sigma I (-\bm{Q}_F)$ and $\Delta I (+\bm{Q}_F)=-\Delta I (-\bm{Q}_F)$.
Moreover, the circular dichroism at both $\pm\bm{Q}_F$ exhibits oscillatory spectra.

We performed simulations of the resonant x-ray scattering spectra using the \texttt{EDRIXS} package~\cite{WANG2019}. 
By incorporating the second-order scattering process, the simulations explicitly account for both the scattering geometry and the polarization state of the scattered x-rays.
The calculated REXS intensity spectra for LCP and RCP x-rays, together with the corresponding circular dichroism and x-ray absorption spectra, are shown in Figs.~\ref{fig:spectra}(a)--\ref{fig:spectra}(d) as dashed lines. 
These spectra are in good agreement with the experimental observations. 
The REXS spectra calculated for representative magnetic structures are presented in Fig.~\ref{fig:calcspectra} in Appendix~\ref{app:spectra}.
Only a dual sinusoidal modulation of spin and AMD moments gives distinct spectral line shapes for LCP and RCP incident x-rays, generating the oscillatory circular dichroism signals. 
The persistence of the dichroism across the multiplet structure of the $L_{2,3}$ edges indicates that it is not associated with a particular intermediate state but reflects the local magnetic symmetry of the underlying magnetic configuration.
Thus, the observed scattering circular dichroism reveals the hidden component, namely, the spatially modulated ferroic AMD at the Fe1 and Fe2 sites.
As illustrated in Fig.~\ref{fig:spectra}(e), the configurations of both $\bm{s}_F$ and the AMD at $\phi = \pi/2$ and $3\pi/2$ are related by a $c$-glide symmetry without time-reversal operation located at $\phi = \pi$.
This symmetry relation accounts for the opposite circular-dichroism intensities observed at $\pm \bm{Q}_F$, where the scattering with LCP and RCP at $\pm \bm{Q}_F$ is related by mirror symmetry.

\section{Conclusion}
The present results show that the scattering circular dichroism from the antiferromagnetic spin cycloid of the multiferroic BiFeO$_3$ originates from spatially modulated AMDs.
When the N\'eel vector $\bm{N}$ is aligned along the [$\bar{1}10$]$_\text{rh}$ direction, the local magnetic symmetry allows an AMD along the [111]$_\text{rh}$ direction.
In contrast, for $\bm{N} \parallel [111]_\text{rh}$, the local magnetic symmetry forbids the magnetic dipole and allows only higher-rank multipoles, such as the magnetic octupole.
This phase-dependent local symmetry allows a spatially modulated ferroic AMD component along the spin cycloid.
The observed dichroism, which cannot be explained by the established cycloidal and Dzyaloshinskii-Moriya-interaction-induced sinusoidal spin structures, arises from the interference between the waves scattered from the modulated AMDs and those from the sinusoidal spin component.
These findings demonstrate that long-wavelength antiferromagnetic textures can host spatially modulated AMDs as hidden magnetic multipoles beyond the spin and orbital moments, and that scattering circular dichroism provides a momentum-resolved probe of such multipoles.
More broadly, they suggest a route toward controlling spatially modulated magnetic multipoles through coupled ferroic degrees of freedom, such as ferroelectric polarization, cycloidal magnetism, orbital anisotropy, and strain.

\begin{acknowledgments}
The authors are grateful to H. Nakao, K. Ono, A. Hashimoto, M. Sakamaki, K. Amemiya, and Y. Kotani for their support of experiments and sample preparation.
This work was supported by JSPS KAKENHI (Grants No. JP19K23590, No. JP19H04399, No. JP20K20107, No. JP23K17145, No. JP24K03205, No. JP24H01685, and No. JP25K03387), by PRESTO (JPMJPR2102) and CREST (JPMJCR1861 and JPMJCR2435), Japan Science and Technology Agency (JST),
and by the MEXT Quantum Leap Flagship Program (MEXT Q-LEAP) Grant No. JPMXS0118068681.
The synchrotron radiation experiments were performed at the Photon Factory with the approval of the Photon Factory Program Advisory Committee (Proposals No. 2012S2-005, No. 2015S2-007, No. 2021S2-004, No. 2023PF-S003, No. 2023G664, No. 2024G615, and No. 2024G630)
and at the NanoTerasu BL14U of the Photon Science Innovation Center (PhoSIC) under the Coalition Program.
\end{acknowledgments}

\section*{Author contributions}
T.I. and S.K. prepared the single crystals.
Y.I., T.H., and Y.Y. fabricated the lamella samples by FIB milling.
Y.I., T.H., T.N., A.S., and Y.Y. performed the experiments and collected the data.
H.N. and Y.Y. performed the micromagnetic simulations using \texttt{MuMax3}, and Y.I., N.S., and Y.Y. simulated the x-ray absorption and resonant x-ray scattering spectra using the \texttt{EDRIXS} Python package.
Y.I., T.A., and Y.Y. discussed the results, wrote the paper, and prepared the figures.
All authors commented on the final version of the manuscript.

\appendix

\section{Scattering formalism for circular dichroism}
\label{app:formalism}
The atomic scattering factor of the $E1$-$E1$ resonant scattering arising from a magnetic dipole is expressed as an antisymmetric matrix,
\begin{equation}
f^{M} = 
\begin{pmatrix}
0 & f^{M}_{z} & -f^{M}_{y} \\
-f^{M}_{z} & 0 & f^{M}_{x} \\
f^{M}_{y} & -f^{M}_{x} & 0
\end{pmatrix}.
\label{eq:fmag}
\end{equation}
Here, we take the $z$ and $x$ axes along the [111]$_\text{rh}$ and [$\bar{1}10$]$_\text{rh}$ directions, respectively, as in Sec.~\ref{sec:symmetry}.
As illustrated in Fig.~\ref{fig:amd}(b), at the cycloidal phases $\phi = \pi/2$ and $3\pi/2$, the canted spin component $\bm{s}_F$ along the $y$ axis induced by the DM interaction aligns ferroically at the Fe1 and Fe2 sites, giving rise to $f^{M}_{y}$.
The AMD induced by the $\bm{\Omega}$ rotation has ferroic $z$ and antiferroic $x$ components.
Consequently, only the $z$ component of the AMD ($t_z$) contributes to the REXS intensity at $\pm\bm{Q}_F$, through $f^{M}_{z}$.
At $\phi = 0$ and $\pi$, by contrast, no ferroic magnetic dipole component is present, and these phases give no contribution to the scattering factors (Appendix~\ref{app:cycloid}).
We denote the scattering factors originating from $\bm{s}_F$ and $t_z$ by $f^{s_F}_{y}$ and $f^{t_z}_{z}$, respectively.

In the Cartesian coordinate system shown in Fig.~\ref{fig:rexs}(a), with the incident x-rays along the $z$ axis and the scattering plane parallel to the $xz$ plane, the $\sigma$ and $\pi$ polarization vectors of the incident and scattered x-rays are given by
\begin{equation}
\begin{aligned}
\bm{\varepsilon}_{\sigma} &= \bm{\varepsilon}^{\prime}_{\sigma} = (0,1,0), \\
\bm{\varepsilon}_{\pi} &= (1,0,0), \\
\bm{\varepsilon}^{\prime}_{\pi} &= (\cos2\theta,0,-\sin2\theta),
\end{aligned}
\label{eq:pol}
\end{equation}
where $2\theta$ denotes the scattering angle.
The structure factor is given by $F(\bm{q}) = \sum_j f^{j}e^{i\bm{q}\cdot\bm{r}_j}$, where $f^{j}$ is the scattering factor matrix of the $j$th Fe ion at the position $\bm{r}_j$ and $\bm{q}$ is the scattering vector.
Its components in the $\sigma$ and $\pi$ polarization bases are $F_{\varepsilon^{\prime}\varepsilon} = \bm{\varepsilon}^{\prime}\cdot F\,\bm{\varepsilon}$.
When $\bm{s}_F$ and $t_z$ are modulated with the propagation vector $\bm{\tau}$, these components are given by
\begin{equation}
\begin{aligned}
F_{\sigma^{\prime}\sigma}(\bm{q}) &= 0, \\
F_{\pi^{\prime}\sigma}(\bm{q}) &= \cos2\theta \sum_j f^{t_z}_{z}\sin(\bm{\tau}\cdot\bm{r}_j)\, e^{i\bm{q}\cdot\bm{r}_j}, \\
F_{\sigma^{\prime}\pi}(\bm{q}) &= -\sum_j f^{t_z}_{z}\sin(\bm{\tau}\cdot\bm{r}_j)\, e^{i\bm{q}\cdot\bm{r}_j}, \\
F_{\pi^{\prime}\pi}(\bm{q}) &= -\sin2\theta \sum_j f^{s_F}_{y}\sin(\bm{\tau}\cdot\bm{r}_j)\, e^{i\bm{q}\cdot\bm{r}_j}.
\end{aligned}
\label{eq:Fcomp}
\end{equation}
The REXS intensity for circular polarization $\chi=\pm 1$ is given by
\begin{equation}
I_{\chi}= \frac{1}{2}\sum_{\varepsilon, \varepsilon^\prime}|F_{\varepsilon^\prime\varepsilon}|^2+\chi\, \text{Im}\left(F_{\sigma^\prime\sigma} F_{\sigma^\prime\pi}^*+F_{\pi^\prime\sigma}F_{\pi^\prime\pi}^*\right).
\label{eq:Ichi}
\end{equation}
Since $F_{\sigma'\sigma}$ vanishes at the magnetic satellite reflections, the circular dichroism $\Delta I = I_{+1} - I_{-1}$ is governed by the interference between $F_{\pi'\sigma}$ and $F_{\pi'\pi}$:
\begin{equation}
\Delta I(\bm{q}) = -\frac{N^2}{2}\, {\rm Im}\left(\cos2\theta\sin2\theta\, f_{z}^{t_z}f_{y}^{s_F*}\right) \delta\left[\bm{q}-(\bm{G}\pm\bm{\tau})\right],
\label{eq:CDfull}
\end{equation}
where $N$ is the number of Fe ions, $\delta(x)$ is the delta function, and $\bm{G}$ is the reciprocal lattice vector.
Equation~(\ref{eq:CDfull}) indicates that the interference between the waves scattered from $\bm{s}_F$ and $t_z$ produces circular dichroism when $f_z^{t_z}$ and $f_y^{s_F}$ have different energy dependences in their real and imaginary parts.
As shown in Fig.~\ref{fig:fxyz}(b), $f_z^{t_z}$ and $f_y^{s_F}$ intrinsically exhibit such different energy dependences, resulting in a finite $\Delta I$.
Furthermore, because of the factor $\sin2\theta$, the circular dichroism has opposite signs for the waves scattered to opposite sides of the incident beam, yielding opposite circular-dichroism intensities at $\pm \bm{Q}_F$.

\section{Strain-induced alignment of magnetic domains}
\label{app:strain}
To examine the influence of uniaxial anisotropy on magnetic domain alignment, we performed micromagnetic simulations using the \texttt{MuMax3} software with varying anisotropy constants $K_u$ (J/m$^3$) along the $y$ direction.
The grid and cell sizes are (1024, 1024, 1) and (5~nm, 5~nm, 1~nm), respectively.
In the simulations, we used a saturation magnetization of $M_s = 6 \times 10^2$~A/m, an exchange stiffness of $A = 1.5 \times 10^{-11}$~J/m, and a Dzyaloshinskii-Moriya interaction constant of $D = 0.5 \times 10^{-2}$~J/m$^2$, originating from the ferroelectric polarization.

\begin{figure*}[tbp]
\begin{center}
\includegraphics[width=17.5cm]{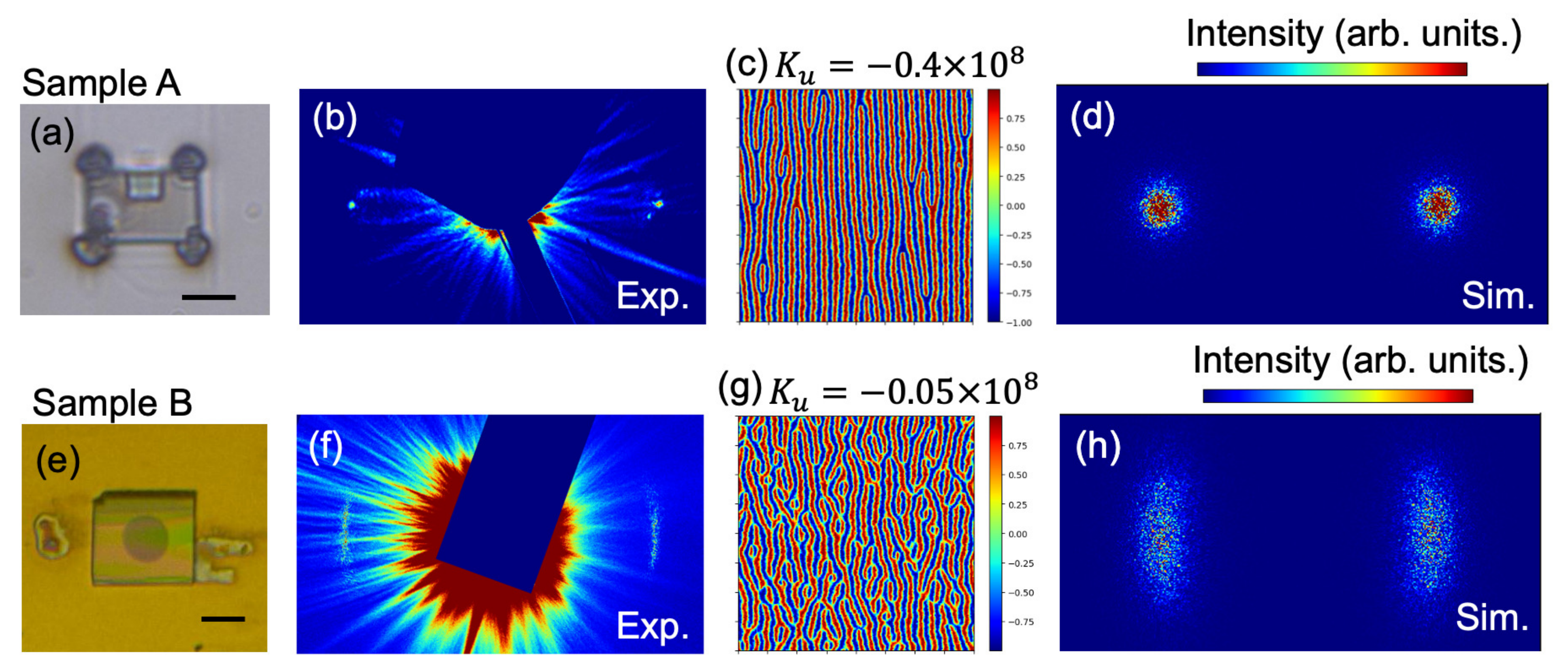}
\caption{
Alignment of magnetic domains induced by uniaxial strain.
(a),(e) SEM images of samples A and B, which were rigidly and softly mounted on the Si$_3$N$_4$ membrane, respectively.
(b),(f) REXS patterns of samples A and B at the Fe $L_3$ edge.
(c),(g) Simulated stripe domain patterns for $K_u = -0.4 \times 10^{8}$~J/m$^3$ and $-0.05 \times 10^{8}$~J/m$^3$, respectively.
(d),(h) Two-dimensional Fourier transforms of (c) and (g).
}
\label{fig:mumax}
\end{center}
\end{figure*}

For weak anisotropy, the stripe domains lack global alignment, and their Fourier transform shows broad intensity along the azimuthal direction.
For strong anisotropy, the stripe domains align along a single direction, and their Fourier transform shows a pair of sharp peaks.
These results reproduce the experimental REXS patterns of the rigidly mounted sample A and the softly mounted sample B in Fig.~\ref{fig:mumax}, indicating that the twofold magnetic scattering in Fig.~\ref{fig:rexs}(b) arises from cycloidal-domain selection by uniaxial strain.

The influence of strain on the AMD-related dichroic response was evaluated as described in Sec.~III of the Supplemental Material~\cite{SM}.
The circular dichroism remains finite even without strain, while the AMD-related scattering amplitude is sensitive to strain.

\section{Calculation of resonant x-ray scattering spectra for representative magnetic structures}
\label{app:spectra}
The XAS, XMCD, and REXS spectra at the Fe $L_{2,3}$ edges were calculated for a single Fe$^{3+}$ ($3d^5$) ion using the \texttt{EDRIXS} package~\cite{WANG2019}.
The Slater integrals were reduced from their Hartree-Fock values to 65\% for $F^2_{dd}$ and $F^4_{dd}$ in the initial state, and to 85\% for $F^2_{dd}$ and $F^4_{dd}$, 95\% for $F^2_{dp}$, and 72.5\% for $G^1_{dp}$ and $G^3_{dp}$ in the intermediate state.
Atomic values were used for the spin-orbit coupling of the Fe $3d$ shell, and that of the Fe $2p$ shell was scaled to 90\% of its atomic value.
The crystal field was calculated in the point-charge model from the positions of the surrounding O ions, including the octahedral rotation $\Omega$ (Sec.~II of the Supplemental Material~\cite{SM}), and scaled to $10Dq = 1.15$~eV.
The molecular field, coupled to $\bm{l}+2\bm{s}$ and applied along the local spin direction at each site, was treated as an adjustable parameter, and a value of 5~meV, which gives the best agreement with the experimental spectra, was used.
The core-hole lifetime broadening was set to 0.6~eV (half width at half maximum), and the spectra were averaged over the ten lowest-energy initial states with Boltzmann weights at 300~K.
The cycloid was represented by the four phases $\phi = 0$, $\pi/2$, $\pi$, and $3\pi/2$.
The remaining parameters were chosen to reproduce the experimental XAS spectra.

\begin{figure*}[tbp]
\begin{center}
\includegraphics[width=14cm]{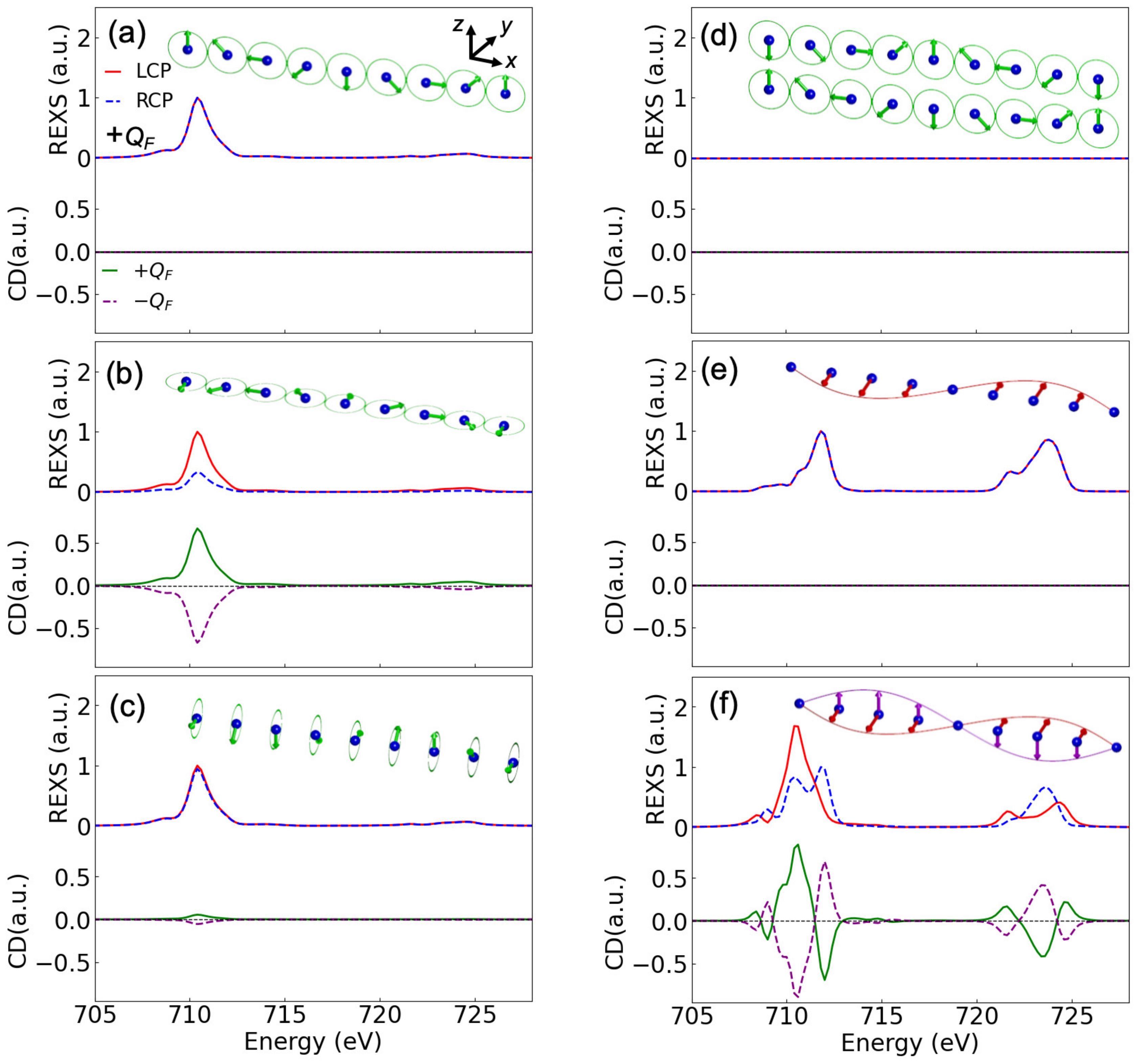}
\caption{
Calculated REXS spectra for representative magnetic structures:
(a) a cycloidal structure with the cycloidal plane within the scattering plane,
(b) a cycloidal structure with the cycloidal plane perpendicular to the scattering plane,
(c) a helical structure,
(d) an antiferromagnetic cycloidal structure,
(e) a sinusoidal modulation of only spin moments, and
(f) a dual sinusoidal modulation of spin and AMD moments.
Red and blue lines represent the REXS intensities at $+\bm{Q}_F$ for LCP and RCP x-rays, respectively.
Green and purple lines represent the circular dichroism ($I_{\rm LCP}-I_{\rm RCP}$) at $+\bm{Q}_F$ and $-\bm{Q}_F$, respectively.
}
\label{fig:calcspectra}
\end{center}
\end{figure*}

We calculated the REXS spectra for representative magnetic structures (Fig.~\ref{fig:calcspectra}).
The antiferromagnetic cycloidal structure, which corresponds to the established magnetic structure of BiFeO$_3$, exhibits no circular dichroism [Fig.~\ref{fig:calcspectra}(d)].
Likewise, the sinusoidal modulation of only spin moments, corresponding to the canted spin component, yields no circular dichroism [Fig.~\ref{fig:calcspectra}(e)].
For comparison, we also calculated the spectra for other noncollinear magnetic structures.
No circular dichroism appears for the cycloidal structure whose cycloidal plane lies within the scattering plane [Fig.~\ref{fig:calcspectra}(a)].
Circular dichroism appears for the cycloidal structure whose cycloidal plane is perpendicular to the scattering plane and for the helical structure [Figs.~\ref{fig:calcspectra}(b) and \ref{fig:calcspectra}(c)].
In these cases, however, the LCP and RCP spectra share the same line shape and differ only in intensity.
Only the dual sinusoidal modulation of spin and AMD moments yields distinct line shapes for LCP and RCP x-rays and the oscillatory circular dichroism observed in BiFeO$_3$ [Fig.~\ref{fig:calcspectra}(f)].

\section{$\Omega$ rotation dependence of the AMD moment}
\label{app:omega}
\begin{figure*}[tbp]
\begin{center}
\includegraphics[width=17.5cm]{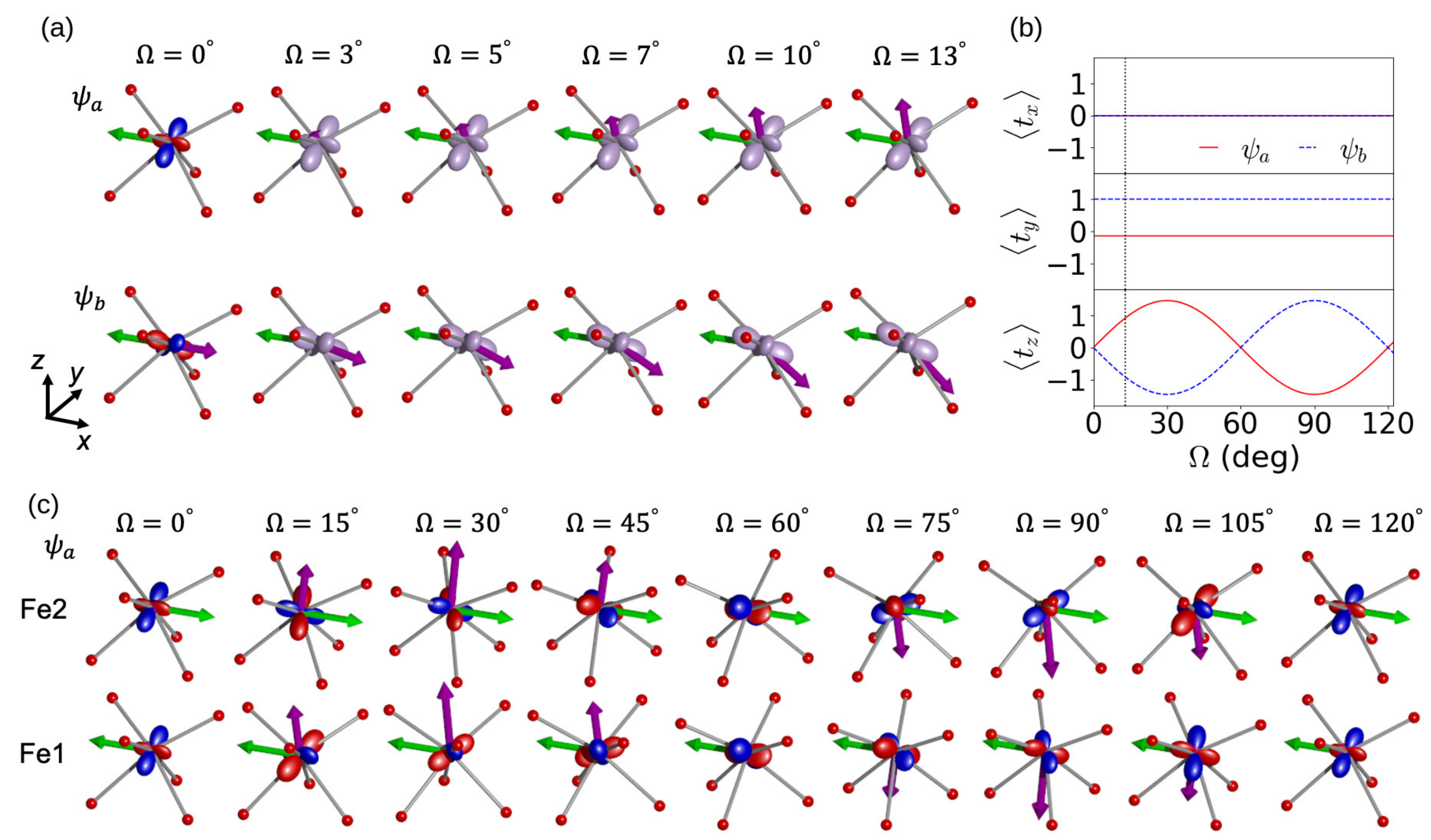}
\caption{
$\Omega$ rotation dependence of the AMD moment at $\phi = \pi/2$.
(a) Spin, orbital, and AMD configurations of the $\psi_a$ (upper panels) and $\psi_b$ (lower panels) states at the Fe1 site for various $\Omega$.
(b) Expectation values of the AMD moments $\langle t_\alpha \rangle$ ($\alpha = x, y, z$) as a function of $\Omega$ for $\psi_a$ (red solid lines) and $\psi_b$ (blue dashed lines).
The vertical dotted line indicates the experimental value $\Omega = 13^{\circ}$.
(c) Spin, orbital, and AMD configurations of the $\psi_a$ states at the Fe1 and Fe2 sites for various $\Omega$.
}
\label{fig:omega}
\end{center}
\end{figure*}

Figure~\ref{fig:omega} shows the dependence of the AMD moments on the octahedral rotation angle $\Omega$ at $\phi = \pi/2$.
As $\Omega$ increases, the anisotropy of the $\psi_a$ and $\psi_b$ orbital states changes, and the $z$ component of the AMD, $t_z$, increases [Fig.~\ref{fig:omega}(a)].
As shown in Fig.~\ref{fig:omega}(b), $t_z$ varies periodically with a period of $120^{\circ}$, reflecting the threefold symmetry of the crystal structure, and has opposite signs for $\psi_a$ and $\psi_b$.
In contrast, $t_x$ takes different values for the two states but is independent of $\Omega$, and $t_y$ vanishes in both states.
At the Fe1 and Fe2 sites, the opposite rotation angles and antiparallel spins lead to the same sign of $t_z$, resulting in a ferroic $t_z$ component [Fig.~\ref{fig:omega}(c)].

\section{Magnetic dipole configuration in the cycloidal structure}
\label{app:cycloid}
\begin{figure*}[tbp]
\begin{center}
\includegraphics[width=15cm]{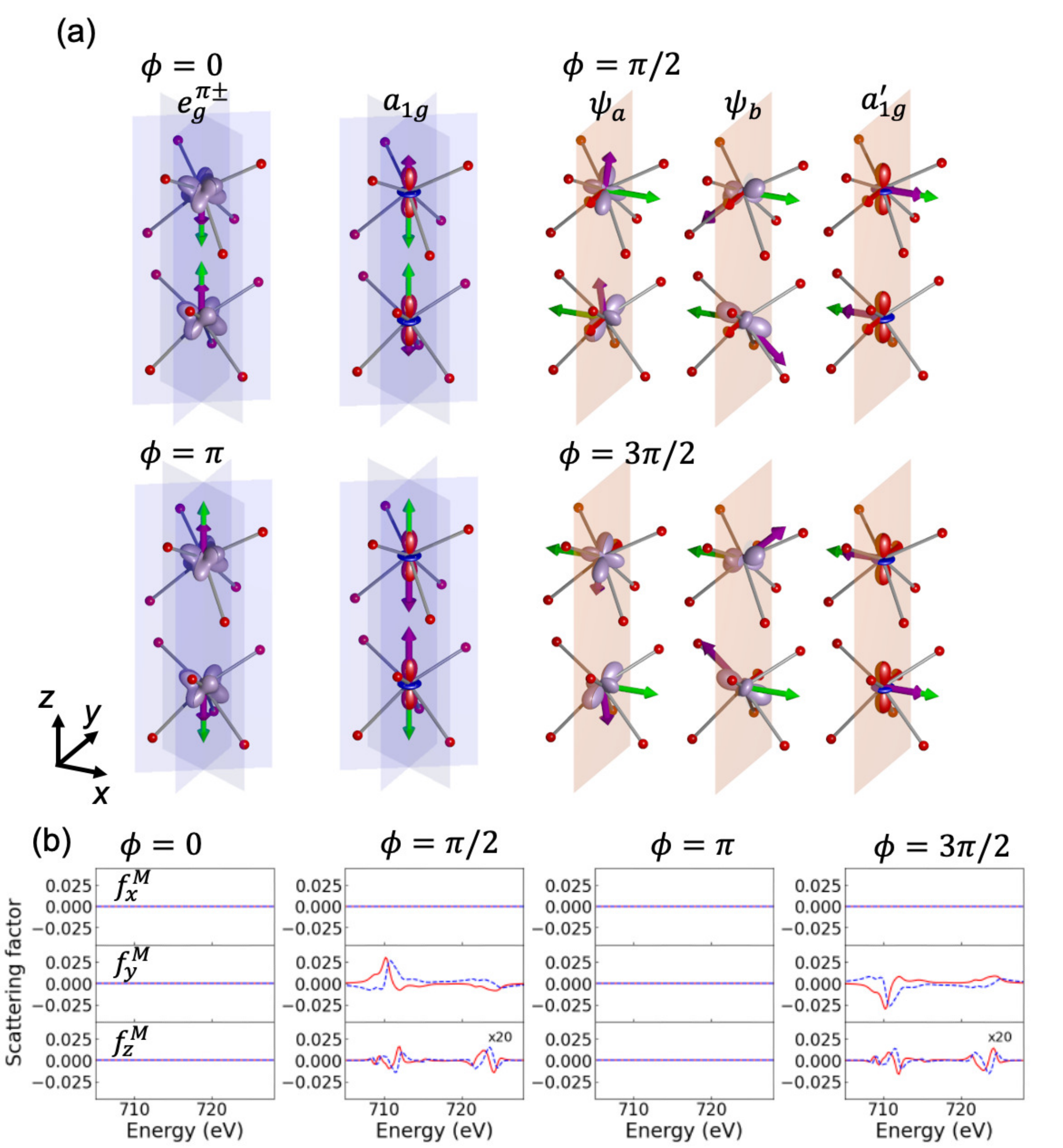}
\caption{
Magnetic dipole configuration in the cycloidal structure.
(a) Spin, orbital, and AMD configurations at the cycloidal phases $\phi = 0$, $\pi/2$, $\pi$, and $3\pi/2$.
(b) Energy dependence of the magnetic scattering factors $f^{M}_{\alpha}$ ($\alpha = x, y, z$) for the Fe1 and Fe2 clusters at each phase.
Red solid and blue dashed lines represent the real and imaginary parts, respectively.
}
\label{fig:fxyz}
\end{center}
\end{figure*}

Figure~\ref{fig:fxyz}(a) shows the spin, orbital, and AMD configurations at the cycloidal phases $\phi = 0$, $\pi/2$, $\pi$, and $3\pi/2$.
At $\phi = 0$ and $\pi$, the AMD is oriented parallel to the spin and therefore aligns antiferroically at the Fe1 and Fe2 sites.
At $\phi = \pi/2$ and $3\pi/2$, the AMD in the $\psi_a$ and $\psi_b$ states has a ferroic $z$ component.
Figure~\ref{fig:fxyz}(b) shows the corresponding magnetic scattering factors $f^{M}_{\alpha}$ ($\alpha = x, y, z$).
At $\phi = \pi/2$ and $3\pi/2$, $f^{M}_{y}$ and $f^{M}_{z}$ arise from the canted ferroic spin component and the AMD, respectively, and exhibit different energy dependences.
At $\phi = 0$ and $\pi$, the scattering factors vanish because all dipoles align antiferroically.

\bibliography{reference,sm_note}

\clearpage
\onecolumngrid

\begin{center}
  {\Large\bfseries Supplemental Material
  for ``Scattering Circular Dichroism from Spatially Modulated
  Anisotropic Magnetic Dipoles in BiFeO$_3$''}\\[0.5em]
\end{center}

\setcounter{section}{0}
\renewcommand{\thesection}{S\arabic{section}}
\setcounter{figure}{0}
\renewcommand{\thefigure}{S\arabic{figure}}
\setcounter{table}{0}
\renewcommand{\thetable}{S\arabic{table}}
\setcounter{equation}{0}
\renewcommand{\theequation}{S\arabic{equation}}

\section{Magnetic scattering in BiFeO$_3$}
\label{sm:extinction}

\begin{figure*}[b]
\begin{center}
\includegraphics[width=17cm]{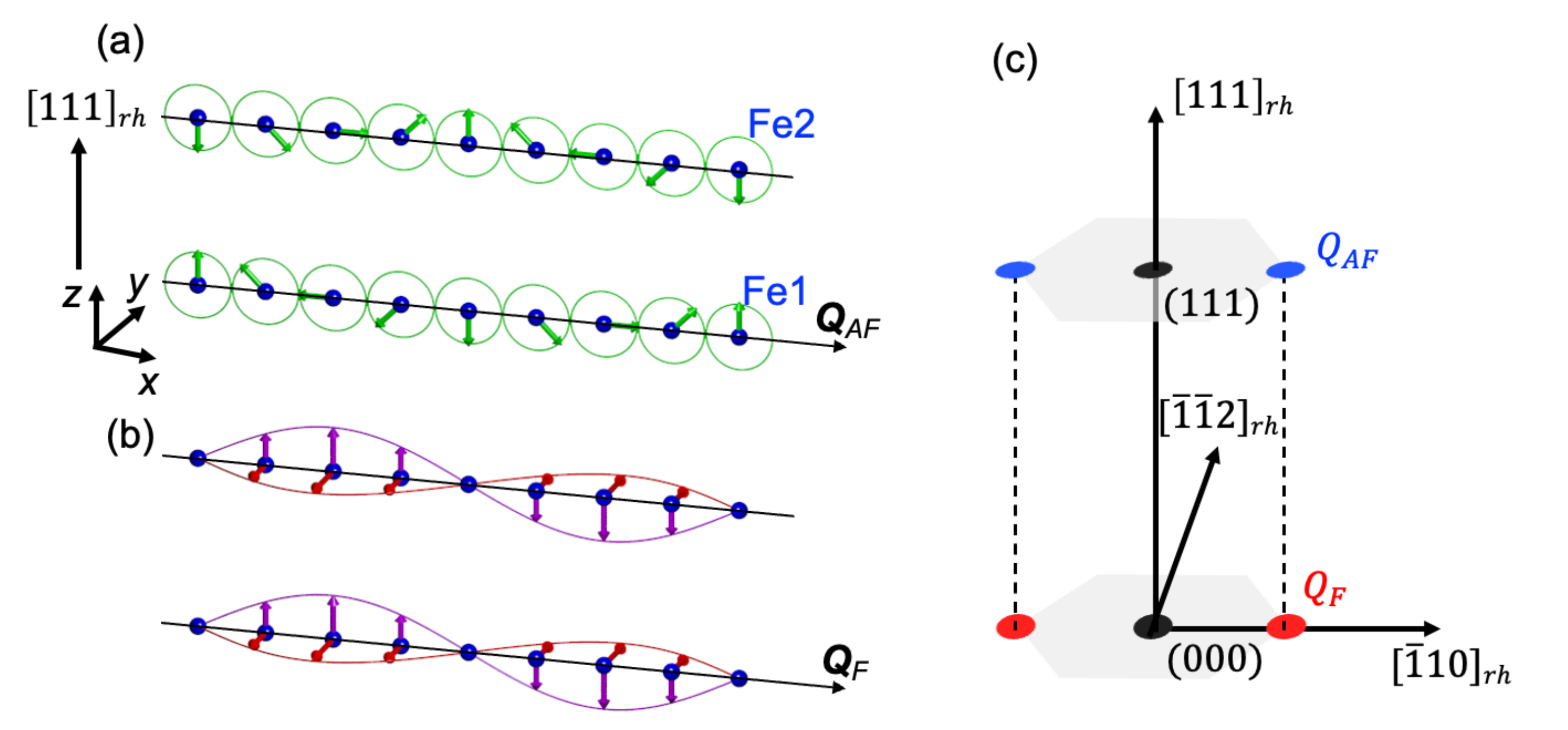}
\caption{
Magnetic scattering peaks in reciprocal space.
(a) Antiferromagnetic cycloidal spin structure.
The spin moments at the Fe1 and Fe2 sites are aligned antiferromagnetically, with the modulation vector $\bm{Q}_{AF} = (1-q_m,1+q_m,1)_{\rm rh}$.
(b) Sinusoidal structures of the canted spin component and the $z$ component of the AMD ($t_z$), represented by red and purple arrows, respectively.
Both components have the same sign at the Fe1 and Fe2 sites and are modulated with the modulation vector $\bm{Q}_F = (-q_m,q_m,0)_{\rm rh}$.
(c) Magnetic peaks in reciprocal space arising from the cycloidal and sinusoidal modulations.
Black, red, and blue circles represent the reciprocal lattice points, $\bm{Q}_F$, and $\bm{Q}_{AF}$ reflections, respectively.
}
\label{fig:smQF}
\end{center}
\end{figure*}

BiFeO$_3$ exhibits a cycloidal spin structure with a wavelength of 62~nm.
The spin moments at the Fe1 and Fe2 sites are aligned antiferromagnetically [Fig.~\ref{fig:smQF}(a)], and this order is characterized by the modulation vector $\bm{Q}_{AF} = (1-q_m,1+q_m,1)_{\rm rh}$.
Figure~\ref{fig:smQF}(c) shows the reciprocal lattice points and the magnetic peaks in reciprocal space.
The antiferromagnetic cycloidal spin structure contributes to the $\bm{Q}_{AF}$ peaks around the $(1\,1\,1)_{\rm rh}$ reciprocal lattice point.
In contrast, the scattering factors associated with the cycloidal spin moments at the Fe1 and Fe2 sites, which are related by the $c$-glide symmetry, have opposite signs and therefore cancel at the $\pm\bm{Q}_F$ reciprocal points.
Consequently, the antiferromagnetic cycloidal component does not contribute to the scattering at $\pm\bm{Q}_F$.

Owing to the Dzyaloshinskii-Moriya (DM) interaction with the $\bm{D}$ vector along the [111]$_{\rm rh}$ direction, the cycloidal spin moments are slightly canted along the [$\bar{1}\bar{1}2$]$_{\rm rh}$ ($y$) direction when they are perpendicular to the $\bm{D}$ vector.
This results in a sinusoidal spin structure in which the canted components have the same sign at the Fe1 and Fe2 sites [Fig.~\ref{fig:smQF}(b)].
In addition, the $z$ component of the anisotropic magnetic dipole (AMD), $t_z$, of the $3d$ electronic states is also sinusoidally modulated, with a ferroic component along the [111]$_{\rm rh}$ direction, as shown in Fig.~\ref{fig:smQF}(b).
Both the canted spin and AMD components are modulated with $\bm{Q}_{F} = (-q_m,q_m,0)_{\rm rh}$.
Consequently, the scattering factors associated with these sinusoidal modulations contribute to the $\pm\bm{Q}_{F}$ peaks observed in the present REXS experiments.

\section{Electronic states under the trigonal crystal field}
\label{sm:cf}

\begin{figure}[t]
\begin{center}
\includegraphics[width=8cm]{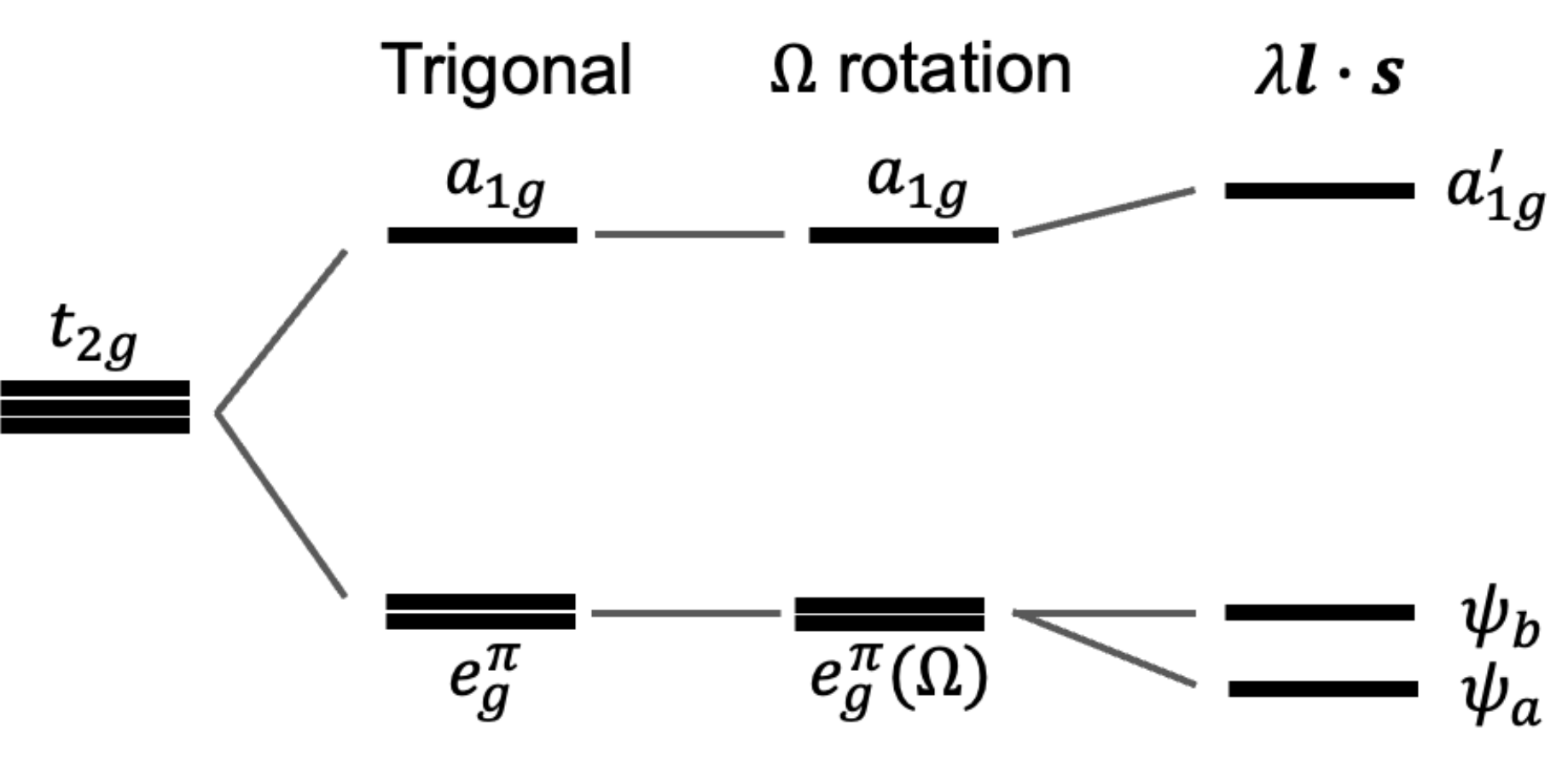}
\caption{Electronic states of the $t_{2g}$ orbitals under the trigonal crystal field and the spin-orbit interaction.}
\label{fig:smt2g}
\end{center}
\end{figure}

In BiFeO$_3$, the Fe$^{3+}$ ions occupy oxygen octahedra that are distorted along the $[111]_{\rm c}$ direction in the pseudocubic (c) notation, which is equivalent to the $[111]_{\rm rh}$ direction.
The local environment of the Fe ion can be described in terms of this trigonal distortion.
In an ideal octahedral ($O_h$) environment, the fivefold degenerate $3d$ orbitals split into doubly degenerate $e_g$ and triply degenerate $t_{2g}$ levels.
The trigonal distortion along the $[111]_{\rm c}$ axis reduces the local symmetry to $C_{3v}$, and the $\bm{\Omega}$ rotation further lowers it to $C_3$.
Consequently, the $t_{2g}$ orbitals split into a nondegenerate $a_{1g}$ state and doubly degenerate $e_g^\pi$ states (Fig.~\ref{fig:smt2g}).

The $d$-orbital wave functions are expressed using the spherical harmonics $Y_{m}^{(2)}(\hat{\bm{r}})$ and the radial function $R_d(r)$ as $\lvert m \rangle = Y_{m}^{(2)}(\hat{\bm{r}}) R_d(r)$, with the quantization axis along the trigonal [111]$_{\rm rh}$ ($z$) axis.
The $a_{1g}$ state corresponds to $\lvert 0 \rangle$, and the $e_g^\pi$ states are given by
\begin{equation}
    |\pi_\pm \rangle =\cos\zeta\,|\pm 2\rangle \mp \sin \zeta\,|\mp 1\rangle,
\end{equation}
where $\zeta$ depends on the trigonal distortion and satisfies $\tan \zeta = 1/\sqrt{2}$ for ideal $O_h$ symmetry.
The $e_g^\pi$ orbitals carry a residual orbital angular momentum,
\begin{equation}
\langle \pi_\pm |\hat{l}_z|\pi_\pm\rangle = \pm (3\cos^2\zeta-1).
\end{equation}
The $\bm{\Omega}$ rotation by the angle $\Omega$ about the $z$ axis modifies the wave functions as
\begin{equation}
    |\pi_\pm \rangle = e^{\mp 2 i\Omega}\cos\zeta\,|\pm 2\rangle \mp e^{\pm i\Omega} \sin \zeta\,|\mp 1\rangle.
\end{equation}

The spin-orbit interaction $H_{LS} = \lambda\,\bm{l}\cdot\bm{s}$ mixes the $|\pi_+\rangle$ and $|\pi_-\rangle$ states and splits them into the $\psi_a$ and $\psi_b$ states, and correspondingly modifies the $a_{1g}$ state into the $a_{1g}^{\prime}$ state (Fig.~\ref{fig:smt2g}).
The states shown in Figs.~3(b), 7, and 8(a) of the main text were obtained by exact diagonalization of the single-electron Hamiltonian $H = H_{\rm CF} + H_{LS} + H_{\rm ex}$ for the $3d$ shell.
The crystal field $H_{\rm CF}$ was calculated in the point-charge model from the positions of the six surrounding O ions, with Fe--O bond lengths of 1.9505 and 2.1126~\AA\ and polar angles of $62.67^\circ$ and $132.12^\circ$ with respect to the [111]$_{\rm rh}$ axis; the azimuthal angles of the O ions are rotated by $+\Omega$ and $-\Omega$ at the Fe1 and Fe2 sites, respectively.
The exchange field $H_{\rm ex} = 2\bm{h}\cdot\bm{s}$ acts on the spin along the local spin direction, with $|\bm{h}| = 0.05$ in units of the crystal-field energy.
The spin-orbit coupling constant was set to $\lambda = 10^{-6}$ in the same units, so that the spin-orbit interaction acts only as an infinitesimal perturbation that selects the orbital states within the degenerate $e_g^\pi$ manifold.
The expectation values $\langle\bm{t}\rangle$ were evaluated for the three lowest-energy states, which correspond to the $\psi_a$, $\psi_b$, and $a_{1g}^{\prime}$ states.

\section{Uniaxial strain effects on the AMD-related dichroism}
\label{sm:strain}

\begin{figure*}[t]
\begin{center}
\includegraphics[width=15cm]{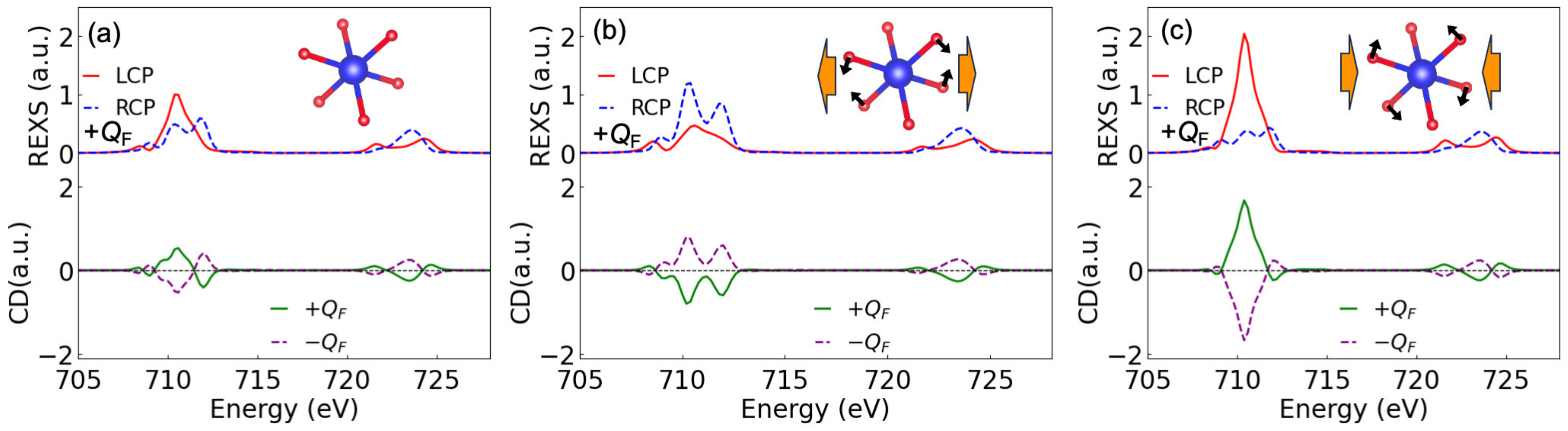}
\caption{
Calculated energy spectra of the REXS and circular-dichroism intensities
(a) without strain and under uniaxial (b) tensile and (c) compressive strain.
Red and blue lines represent the REXS intensities at $+{\bm Q}_F$ for left- and right-circularly polarized (LCP and RCP) incident x-rays, respectively.
Green and purple lines represent the circular dichroism ($I_{\rm LCP}-I_{\rm RCP}$) at $+\bm{Q}_F$ and $-\bm{Q}_F$, respectively.
The insets schematically illustrate the rotations of the O ions associated with the respective strain states.
The rotation angles of the O ions are set to $\pm 0.5^\circ$, corresponding to strains of $\pm 1\%$.
}
\label{fig:smstrain}
\end{center}
\end{figure*}

The strain in the present sample is considered to act primarily as a domain-selection field for the cycloidal propagation direction.
At the same time, since the AMD originates from the local orbital anisotropy, strain can in principle modify the AMD amplitude and the relative weights of the circular-dichroic scattering channels.
To examine the role of strain in the observed dichroic response, we performed spectral simulations including strain-induced modifications of the local electronic anisotropy under both compressive and tensile uniaxial strain.
As shown in Fig.~\ref{fig:smstrain}(a), the circular dichroism remains finite even in the absence of strain, indicating that strain is not the primary origin of the dichroic signal.
Rather, strain modifies the AMD-related scattering response by changing the local orbital anisotropy and, consequently, the interference between the AMD and the DM-induced canted spin components.
Depending on the sign of strain, this can enhance or reduce the circular-dichroic intensity, as shown in Figs.~\ref{fig:smstrain}(b) and \ref{fig:smstrain}(c).
Thus, in the present sample, strain plays two roles: it selects the cycloidal propagation domain, giving rise to the twofold magnetic scattering pattern, and it modifies the circular-dichroic response associated with the AMD.
The observed dichroism should therefore be regarded as a strain-modified, but not strain-induced, resonant response of the AMD in BiFeO$_3$.

\end{document}